\documentstyle[galley,epsf,epsfig]{mn2e}

\title[NITROGEN ENRICHMENT IN ATMOSPHERES OF A- AND F-TYPE SUPERGIANTS]
{NITROGEN ENRICHMENT IN ATMOSPHERES OF A- AND F-TYPE SUPERGIANTS}

\author[L.S. Lyubimkov et al.]
{Leonid~S.~Lyubimkov,$^1$\thanks{E-mail: lyub@crao.crimea.ua (LSL); dll@astro.as.utexas.edu (DLL); serkor@skyline.od.ua (SAK)} David~L.~Lambert,$^2$$^\star$ Sergey~A.~Korotin,$^3$$^\star$ \and
Dmitry~B.~Poklad,$^1$ Tamara~M.~Rachkovskaya$^1$ and Sergey~I.~Rostopchin$^2$\\
$^1$Crimean Astrophysical Observatory, Ukraine\\
$^2$The W.J. McDonald Observatory, The University of Texas at Austin, USA\\
$^{3}$Astronomical Observatory of the Odessa National University, Ukraine\\}

\begin{document}

\date{Accepted. Received ; in original form}

\pagerange{\pageref{firstpage}--\pageref{lastpage}} \pubyear{}

\maketitle

\label{firstpage}

\begin{abstract}
Using new accurate fundamental parameters of 30 Galactic A and F supergiants, namely their effective temperatures $T_{\rm eff}$ and surface gravities $\log g$, we implemented a non-LTE analysis of the nitrogen abundance in their atmospheres. It is shown that the non-LTE corrections to the N abundances increase with $T_{\rm eff}$. The nitrogen overabundance as a general feature of this type of stars is confirmed. A majority of the stars has a nitrogen excess $[N/Fe]$ between 0.2 and 0.9 dex with the maximum position of the star's distribution on $[N/Fe]$ between 0.4 and 0.7 dex. The N excesses are discussed in light of predictions for B-type main sequence (MS) stars with rotationally induced mixing and for their next evolutionary phase, i.e. A- and F-type supergiants that have experienced the first dredge-up. Rotationally induced mixing in the MS progenitors of the supergiants may be a significant cause of the nitrogen excesses. When comparing our results with predictions of the theory developed for stars with the mixing, we find that the bulk of the supergiants (28 of 30) show the N enrichment that can be expected (i) either after the MS phase for stars with the initial rotational velocities $v_0$ = 200-400 km s$^{-1}$, (ii) or after the first dredge-up for stars with $v_0$ = 50-400 km s$^{-1}$. The latter possibility is preferred on account of the longer lifetime for stars on red-blue loops following the first dredge-up. Two supergiants without a discernible N enrichment, namely HR 825 and HR 7876, may be post-MS objects with the relatively low initial rotational velocity of about 100 km s$^{-1}$. The suggested range for $v_0$ is approximately consistent with inferences from the observed projected rotational velocities of B-type MS stars, progenitors of A and F supergiants. 
\end{abstract}

\begin{keywords}
stars: abundances - stars: evolution - supergiants
\end{keywords}

\section{Introduction}

      Nitrogen is one of the key light chemical elements, whose atmospheric abundances can significantly change during the stellar evolution. Therefore, nitrogen provides an opportunity for testing modern evolutionary theories. We consider here stars with masses $M$ between 4 and 20 $M_\odot$, which are observed at first as early and middle B-type main sequence stars and next as A, F and G-type supergiants. The latter objects are the focus of the present paper.        

      Luck \& Lambert (1985) showed that non-variable F, G, and early K supergiants reveal a systematic nitrogen overabundance that is accompanied with a systematic carbon underabundance (carbon is another key element). These data showed that there is an anticorrelation between the N and C abundances (see, e.g., Lyubimkov's 1998 review). These N and C anomalies were considered to be a result of deep mixing during the red giant/supergiant evolutionary phase (known as the first dredge-up), which leads to transfer of the CNO-cycle products from stellar interiors to their surfaces. 

      Results for early F-type supergiants drawn from N I lines are sensitive to non-LTE effects, which Luck and Lambert modelled with a simple model atom for nitrogen. (LTE is local thermodynamic equilibrium). One may mention as well LTE analyses of the N and C abundances performed by Luck \& Wepfer (1995) for 38 F- and G-type bright giants (luminosity class II), as well as by Smiljanic et al. (2006) for 15 supergiants of types K3-A8. These authors found that the N excess up to 0.7-0.9 dex and the C deficiency down to -0.6 dex are typical for such stars.  It is interesting that a similar spread of the N excess, namely from zero to 0.9 dex, is found by us for our programme supergiants (see below).
    
      A refined non-LTE analysis of N I lines for late-B through F supergiants implemented by Takeda \& Takada-Hidai (1995) confirmed that the atmospheric N abundances for these stars tend to be significantly enhanced as compared with the solar abundance. A non-LTE approach was applied as well by Venn (1995) in the N and C abundance analysis for a sample of A supergiants. Later, Venn \& Przybilla (2003) updated the non-LTE nitrogen abundances for these stars using new atomic data and confirmed that N was overabundant. More recently, Przybilla et al. (2006) and Schiller \& Przybilla (2008) reported non-LTE nitrogen abundances for five A- and B-type supergiants with all showing a considerable nitrogen overabundance. Thus, the 1985 conclusion on the systematic N excess in F, G, and early-K supergiants is confirmed and extended to the A-type supergiants.

      When discussing possible sources of errors in the derived abundances, one should note the important role of the accuracy of two fundamental stellar parameters, namely the effective temperature $T_{\rm eff}$  and surface gravity $g$ ($\log g$ is usually quoted). 
Recent inspection by Lyubimkov et al. (2009, 2010) of current data on $T_{\rm eff}$ and $\log g$ for AFG supergiants revealed that there is a significant scatter between estimates by various authors, even for bright stars. In particular, the difference in $T_{\rm eff}$ for A supergiants attains up to 2400 K, whereas the discrepancy in $\log g$ for F supergiants reaches up to 1.4 dex. Such uncertainties in $T_{\rm eff}$ and $\log g$ can lead to appreciable errors in the derived abundances.
      
      According to Lyubimkov et al. (2010, hereinafter Paper I), a significant improvement in the accuracy of the $\log g$ values can be obtained through application of van Leeuwen's (2007) new reduction of the {\it Hipparcos} parallaxes. Using these new parallaxes in an analysis of 63 Galactic supergiants, they obtained $\log g$ with high accuracy: the typical error in $\log g$ is $\pm$0.06 dex for supergiants with distances $d < 300$ pc and $\pm$0.12 dex for supergiants with  $d$ between 300 and 700 pc. In order to check the derived $T_{\rm eff}$ values, a comparison with determinations using the infrared flux method was used; the mean error of $\pm$120 K in $T_{\rm eff}$ for stars with  $d < 700$ pc (48 in all) was found. For more distant supergiants with  $d > 700$ pc (15 in all), where parallaxes are uncertain or unmeasurable, the typical errors in $\log g$ are 0.2-0.3 dex. Thus, for a rather large sample of AFG supergiants there is an opportunity now to increase markedly the accuracy of determinations of their chemical composition.
       
      We present here our results of a non-LTE nitrogen abundance determination for 8 A-type and 22 F-type supergiants. We used in this analysis the parameters found for the stars in Paper I. The derived N abundances are compared for common stars with above-mentioned data of Takeda \& Takada-Hidai (1995) and Venn \& Przybilla (2003), hereinafter TT'95 and VP'03, respectively. Moreover, we compare our N excesses with data of Przybilla et al. (2006) and Schiller \& Przybilla (2008) for five supergiants, which span an effective temperature $T_{\rm eff}$ and mass $M$ range that overlaps our range but extend mostly to higher $T_{\rm eff}$ and $M$ values. Note that unlike TT'95 and VP'03 the latter two works contain no common stars with our list. Finally, a comparison of the supergiant's observed N enrichment with theoretical expectations is presented.

\section{List of N I lines and the equivalent width measurements}

      High-resolution spectral observations of programme stars were obtained at the McDonald Observatory of the University of Texas at Austin (see Paper I for details). Since N I lines in the visual spectral region are very weak, we used, like other authors, the N I lines in the near infrared region between 7400 and 8730 \AA. These lines are rather strong in spectra of A and F supergiants, but too weak and blended in spectra of cooler G supergiants. Therefore, we consider in this paper only A and F supergiants. (Note that the nitrogen abundance in G supergiants can be inferred from a combination of C I and CN molecular lines, see, e.g., Luck \& Lambert 1985).
\begin{table}
 \centering
 \begin{minipage}{52mm}
  \caption{List of the used N I lines}
  \begin{tabular}{ccc}

 \hline
  Line    & Ex. potential,  & $\log gf$ \\
          & eV              &\\
 \hline
 7423.64  &       10.326    &  -0.705 \\
 7442.30  &       10.330    &  -0.383 \\
 7468.31  &       10.336    &  -0.188 \\
 8184.86  &       10.330    &  -0.285 \\
 8188.01  &       10.326    &  -0.291 \\
 8200.36  &       10.326    &  -0.999 \\
 8210.71  &       10.330    &  -0.708 \\
 8216.33  &       10.336    &   0.133 \\
 8223.13  &       10.330    &  -0.270 \\
 8680.29  &       10.336    &   0.346 \\
 8683.40  &       10.330    &   0.086 \\
 8686.15  &       10.326    &  -0.304 \\
 8703.25  &       10.326    &  -0.321 \\
 8711.71  &       10.330    &  -0.233 \\
 8718.83  &       10.336    &  -0.334 \\
 8728.90  &       10.330    &  -1.064 \\
 \hline
  \end{tabular}
 \end{minipage}
\end{table}

      The N I lines studied in the present work are listed in Table 1. We provide there the line wavelengths in \AA, excitation potentials in eV and oscillator strengths $gf$ according to the VALD database (Kupka et al. 1999, Heiter et al. 2008). It should be noted that there is no marked systematic difference between $gf$-values presented in Table 1 and used by TT'95 and VP'03. Moreover, there is an excellent agreement (within $\pm$0.002 dex) for common lines with the $gf$-values used by Maiorca et al. (2009) in their analysis of the solar N abundance.

      It should be noted that N I lines in the region 8184-8223 \AA\ can be blended by telluric lines. We used only those N I lines whose profiles in observed spectra were clearly separated from telluric lines. So, the N I lines in spectra of the stars with relatively rapid rotation, where such a separation was impossible, have been not considered.

      Equivalent widths $W$ of the N I lines are measured by a Gaussian fitting procedure or by the direct integration. The measured $W$ values for the most of programme stars are presented in Appendix (Table A1). It should be noted that for a part of the stars only profiles of the N I lines were used for the N abundance determination (see below). So, the $W$ values for these stars are absent in Table A1. Note as well that the NI lines 8680.28, 8683.40 and 8686.14 \AA\ are located in the wing of the $P_{13}$ line at 8665.0 \AA.  We took into consideration the $P_{13}$ line wing, when the N abundances from profiles of these three NI lines were derived with the help of synthetic spectra. Their equivalent widths $W$ in Table A1 (in brackets) have been measured with reference to the $P_{13}$ wing as a pseudo-continuum; these Ws were used only for the preliminary N abundance evaluation.

\begin{figure}
\epsfxsize=8truecm
\epsffile{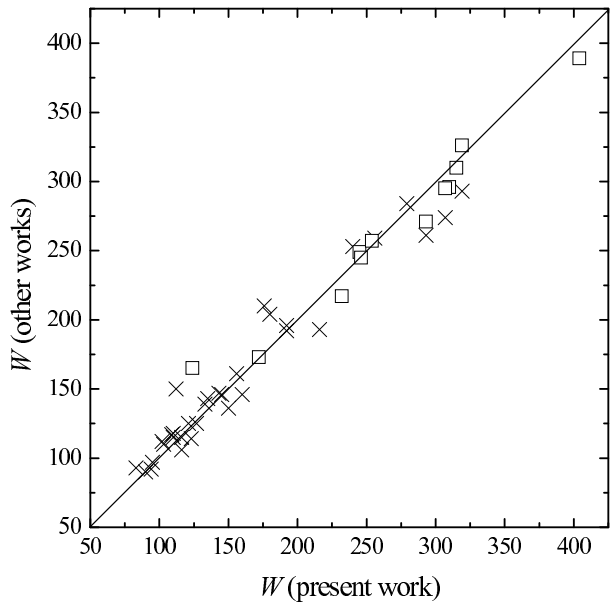}
\caption{Comparison of our equivalent widths of N I lines with data of TT'95 (squares) and VP'03 (crosses).}
\end{figure}
      
      Since we compare below our results on the N abundances for common stars with data of TT'95 and VP'03, it is useful to compare at first our equivalent widths with $W$s from these works. Note that the $W$ values used by VP'03 have been taken from Venn (1995). There are 5 common stars in our and TT'95 lists and 10 stars in our and VP'03 lists (we considered there only stars with temperatures $T_{\rm eff} < 10000$~K). Fig.1 shows the comparison with equivalent widths $W$ used in TT'95 and VP'03 for these stars (squares and crosses, respectively). One may see that equivalent widths of common N I lines are in rather good agreement; no systematic discrepancy exists.

\section{Non-LTE computations of N I lines}

      We used the code MULTI (Carlsson, 1986) as modified by Korotin et al. (1999) for the non-LTE computation of the N I lines.
Since we apply model atmospheres calculated with the code ATLAS9 (Kurucz 1992), the opacity sources for MULTI are taken from ATLAS9, too. In the modified MULTI, the mean intensities (they are needed to obtain the radiative photoionization rates) are calculated for a set of frequencies at each atmospheric layer; then they are stored in a separate block, from where they can be interpolated. Castelli's (2005) tables for the opacity distribution function (ODF) are utilized there. So, we may take into account absorption by the great number of lines, especially in the UV region, that are of importance for photoionization rate calculations. Note that the microturbulent parameter $V_t$ was included in the statistical equilibrium calculations.

\begin{figure}
\epsfxsize=8.5truecm
\epsffile{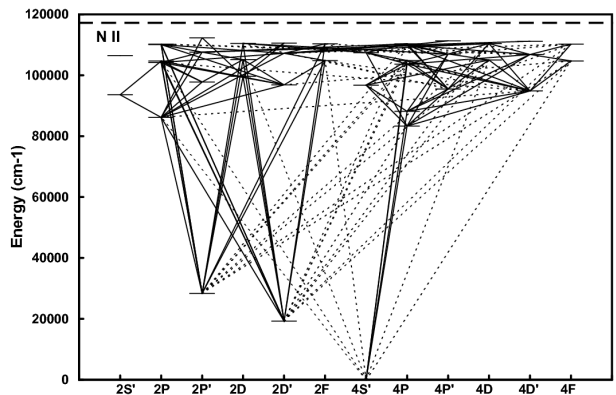}
\caption{Model nitrogen atom.}
\end{figure}

      The adopted N model atom consists of 39 N I levels, 5 N II levels and the ground state of N III. Moreover, additional atomic levels with LTE populations were included in the equation of the particle number conservation, namely 66 N I levels, 38 N II levels and 4 N III levels. Excitation energies were taken for N I terms from Moore (1975) and for N II terms from Eriksson (1983). Photoionization rates for all included levels were calculated with the help of the photoionization cross-sections from OP TopBASE (Cunto \& Mendoza, 1992; Cunto et al. 1993). We selected for detailed consideration 141 radiative bound-bound transitions, both permitted and forbidden. Radiative rates of other 232 very weak transitions were adopted to be fixed and calculated in the LTE approximation. Oscillator strengths for the transitions in question were taken form Hirata \& Horaguchi's (1994) catalogue. Fig.2 shows schematically the used N I model atom; only those transitions are presented there, which were included in the detailed consideration. 
      
      Electron collisional excitation rates were calculated with Frost et al.'s (1998) cross-sections.
Moreover, data from the database ADAS (Summers 2004) were applied as well. For allowed transitions, where upper atomic levels have $n \geq 5$, the formula of van Regemorter (1962) was used. Collisional rates for forbidden transitions between levels with the similar combinations of the principal quantum number $n$ were calculated by the semi-empirical Allen's (1976) formula with a collisional force that equals to 1.

The electron collisional ionization was taken into consideration by the formula of Seaton (1962) with cross-sections on the ionization threshold from OP TopBASE. Collisions with hydrogen atoms for permitted transitions were introduced according to Steenbock \& Holweger (1984); for forbidden transitions Takeda's (1991) method was applied. However, it should be noted that the role of collisions with H atoms seems to be insignificant.

 \begin{table}
 \centering
 \begin{minipage}{85mm}
 \caption{The $\log\epsilon$(N) reanalysis for seven common stars from VP'03 with their parameters $T_{\rm eff}$, $\log g$ and $V_t$ and their equivalent widths of N I lines}
\begin{tabular}{@{}r@{\hspace{0.1cm}}c@{\hspace{0.1cm}}c@{\hspace{0.2cm}}c@{\hspace{0.1cm}}c@{\hspace{0.1cm}}c@{\hspace{0.1cm}}c@{\hspace{0.2cm}}c@{}}
 \hline
   HR  &    HD  & $T_{\rm eff}$, K & $\log g$ & $V_t$,    &  Number  & $\log\epsilon$(N) & $\log\epsilon$(N)\\
       &        &                  &          &km s$^{-1}$& of lines &      (VPÒ03)      &       (ours)     \\
 \hline
  292  &   6130 &   7400           &   1.5    &   4       &    9     &   8.00$\pm$0.14   &   8.02$\pm$0.06  \\
  1242 &  25291 &   7600           &   1.5    &   4       &    7     &   7.96$\pm$0.11   &   7.94$\pm$0.10  \\
  1865 &  36673 &   7400           &   1.1    &   4       &    1     &        8.19       &        8.22      \\
  2839 &  58585 &   8000           &   1.8    &   4       &    3     &   8.26$\pm$0.16   &   8.22$\pm$0.19  \\
  3183 &  67456 &   8300           &   2.5    &   3       &    2     &   8.19$\pm$0.02   &   8.15$\pm$0.01  \\
  6144 & 148743 &   7800           &  1.15    &   8       &    3     &   8.19$\pm$0.13   &   8.20$\pm$0.13  \\
  7876 & 196379 &   7500           &   1.6    &   5       &   11     &   7.61$\pm$0.20   &   7.59$\pm$0.17  \\
 \hline
 \end{tabular}
 \end{minipage}
 \end{table}

\begin{figure}
\epsfxsize=8.5truecm
\epsffile{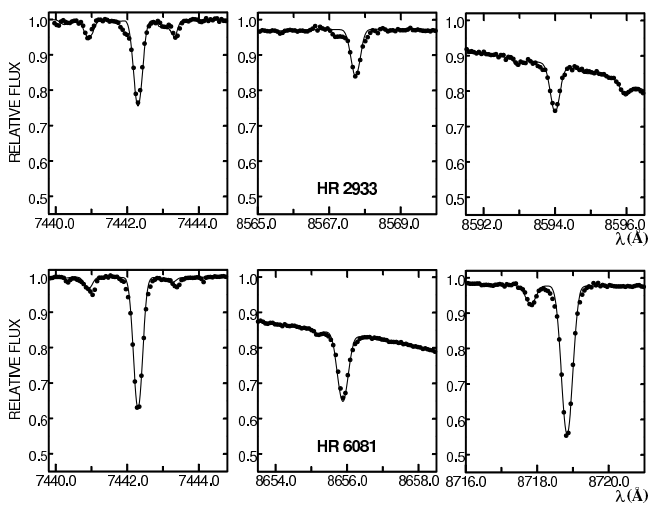}
\caption{The observed and computed profiles of some N~I lines for the stars HR~2933 and HR~6081.}
\end{figure}

      The problem of line blending is important for some N I lines. Using the modified code MULTI we calculated $b$-factors for N I levels and then used them in the synthetic spectra calculations with the code STARSP (Tsymbal 1996). These calculations included all spectral lines from the VALD database in a region of interest. The LTE approach was applied for lines other than the N I lines. Abundances of corresponding elements were adopted in accordance with the $[Fe/H]$ value for each star. Such a method, for instance, was utilized, when the blending of the N I 8680.28 line by the S I 8680.47 line was taken into consideration. Note that the N I 8680.28 line was measured only for 6 programme stars, namely for five A5 stars and one F0 star (Table A1). We evaluated a contribution of the SI line to the equivalent width of the NI+SI blend and found that the S I line contributes only 3-8 per cent in the case of A5 stars and 20 per cent in the case of F0 star.

      The N I 8680.28 line, as well as the N I 8683.40 and 8686.14 lines are located in the wing of the $P_{13}$ line at 8665.0 \AA.  We took into consideration the $P_{13}$ wing, when synthetic spectra were calculated and the N abundances from profiles of these three N I lines were derived. We found no systematic discrepancy in $\log\epsilon$(N) for these lines in comparison with other NI lines and with the mean $\log\epsilon$(N) values; on the contrary, very good agreement exists (see Table A1 and Table 3).

      Supergiants with effective temperatures $T_{\rm eff}$ = 5600-8600 K are considered by us here, so our non-LTE technique is not intended for a study of hotter stars with temperatures, say, $T_{\rm eff} \geq 9000$ K. Since nitrogen in atmospheres of such supergiants is mostly neutral, we included the few number of the N II levels and transitions; consequently, our model atom is simpler than that adopted by VP'03 (see also Przybilla \& Butler 2001 for details). So, it would be interesting to compare the accuracy of our non-LTE technique with that of VP'03. Using our technique, we redetermined the N abundances for common stars from VP'03 using their data, i.e. their parameters $T_{\rm eff}$, $\log g$ and $V_t$ and their equivalent widths $W$ of N I lines (note that the same oscillator strengths $\log gf$ of the lines were applied by us and VP'03). Their $W$ values have been published by Venn (1995); however, not all these $W$s were used in VP'03. But, we could reanalyse seven of the ten common stars, where it was possible to indicate definitely the N I lines used in VP'03. Data of VP'03 for these stars are presented in Table 2; results of our reanalysis are shown in the last column. One may see from Table 2 (two last columns) that the very good agreement with the VP'03 mean N abundances takes place. The difference varies from -0.04 to +0.03 dex. So, we reproduced virtually the results of VP'03. Therefore, we arrive at the following conclusions:\\
(i) the accuracy of our technique for supergiants with $T_{\rm eff}$ = 5600-8600 K is not lower than that of VP'03;\\
(ii) discrepancies in the N abundances between VP'03 and our results (see below) are explained mostly by differences in parameters $T_{\rm eff}$, $\log g$ and $V_t$. Small remaining discrepancies within $\pm$0.04 dex, if they are real, can be explained by small differences in the measured $W$ values and differences in the used non-LTE techniques.

      It should be noted that the N I lines listed in Table 1 are formed from the same lower term. In order to check our non-LTE technique, we also analyzed for a few stars the N I lines formed from other terms. In Fig.3 we display an example of such analysis for two stars with different effective temperatures, namely HR 2933 ($T_{\rm eff}$ = 6690 K) and HR 6081 ($T_{\rm eff}$ = 8370 K). Apart from the lines 7442.30 and 8718.83 from Table 1, the lines 8567.74, 8594.00 and 8655.87 are also presented there (the latter two are in the $P_{14}$ and $P_{13}$ wings, respectively). Synthetic spectra (solid curves) show good agreement with observed profiles (dots) for a single $\log\epsilon$(N) values (the means from Table 3), specifically $\log\epsilon$(N) = 8.40 for HR 2933 and $\log\epsilon$(N) = 8.30 for HR 6081. We may conclude that Fig.3 confirms once again the good accuracy of our non-LTE analysis.

\begin{table*}
 \centering
 \begin{minipage}{180mm}
  \caption{Basic parameters and the nitrogen abundances}
  \begin{tabular}{ccccccccccccc}
\hline
 HR & HD & Name & Sp & $T_{\rm eff}$, & $\log g$ & $V_t$, & $d$, pc & $M/M_\odot$ & $\log\epsilon($Fe) & $\log\epsilon$(N) & $[N/H]$ & $[N/Fe]$\\
& & & & K & & km s$^{-1}$ & & & & & &\\
\hline
  27  &   571  &  22 And       &  F5 II   &  6270  & 2.10  &  3.6  &  380  &  6.1 & 7.41$\pm$0.05 & 8.30$\pm$0.18 & 0.47 &  0.56$\pm$0.19 \\
  292 &  6130  &               &  F0 II   &  6880  & 2.05  &  2.7  &  613  &  7.1 & 7.55$\pm$0.10 & 8.49$\pm$0.12 & 0.66 &  0.61$\pm$0.16 \\
 1017 &  20902 & $\alpha$ Per  &  F5 Ib   &  6350  & 1.90  &  5.3  &  156  &  7.3 & 7.43$\pm$0.09 & 8.41$\pm$0.10 & 0.58 &  0.65$\pm$0.13 \\
 1135 &  23230 & $\upsilon$ Per&  F5 II   &  6560  & 2.44  &  3.5  &  170  &  4.8 & 7.56$\pm$0.06 & 8.24$\pm$0.15 & 0.41 &  0.35$\pm$0.16 \\
 1242 &  25291 &               &  F0 II   &  6815  & 1.87  &  3.2  &  629  &  8.3 & 7.43$\pm$0.11 & 8.34$\pm$0.10 & 0.51 &  0.58$\pm$0.15 \\
 1740 &  34578 &  19 Aur       &  A5 II   &  8300  & 2.10  &  4.3  &  637  &  8.8 & 7.42$\pm$0.12 & 8.12$\pm$0.07 & 0.29 &  0.37$\pm$0.14 \\
 1865 &  36673 & $\alpha$ Lep  &  F0 Ib   &  6850  & 1.34  &  3.9  &  680  & 13.9 & 7.53$\pm$0.08 & 8.73$\pm$0.09 & 0.90 &  0.87$\pm$0.12 \\
 2693 &  54605 & $\delta$ CMa  &  F8 Ia   &  5850  & 1.00  &  7.0  &  495  & 14.9 & 7.51$\pm$0.09 & 8.45$\pm$0.16 & 0.62 &  0.61$\pm$0.18 \\
 3073 &  64238 &  10 Pup       &  F1 Ia   &  6670  & 2.61  &  3.5  &  338  &  4.2 & 7.60$\pm$0.06 & 8.45$\pm$0.17 & 0.62 &  0.52$\pm$0.18 \\
 3102 &  65228 &  11 Pup       &  F7 II   &  5690  & 2.17  &  3.7  &  161  &  5.1 & 7.61$\pm$0.07 & 8.35$\pm$0.25 & 0.52 &  0.41$\pm$0.26 \\
 3183 &  67456 &               &  A5 II   &  8530  & 2.67  &  3.5  &  481  &  5.4 & 7.54$\pm$0.12 & 8.12$\pm$0.04 & 0.29 &  0.25$\pm$0.13 \\
 6081 & 147084 &   $o$ Sco     &  A5 II   &  8370  & 2.12  &  2.8  &  269  &  8.7 & 7.53$\pm$0.17 & 8.30$\pm$0.07 & 0.47 &  0.44$\pm$0.18 \\
 6978 & 171635 &  45 Dra       &  F7 Ib   &  6000  & 1.70  &  4.6  &  649  &  8.2 & 7.41$\pm$0.08 & 8.25$\pm$0.09 & 0.42 &  0.51$\pm$0.12 \\
 7264 & 178524 &  $\pi$ Sgr    &  F2 II   &  6590  & 2.21  &  3.2  &  156  &  5.9 & 7.33$\pm$0.09 & 8.30$\pm$0.12 & 0.47 &  0.64$\pm$0.15 \\
 7796 & 194093 & $\gamma$ Cyg  &  F8 Ib   &  5790  & 1.02  &  5.2  &  562  & 14.5 & 7.46$\pm$0.06 & 8.28$\pm$0.13 & 0.45 &  0.49$\pm$0.14 \\
 7834 & 195295 &  41 Cyg       &  F5 II   &  6570  & 2.32  &  3.6  &  235  &  5.3 & 7.50$\pm$0.07 & 8.20$\pm$0.11 & 0.37 &  0.37$\pm$0.13 \\
\hline
  825 &  17378 &               &  A5 Ia   &  8570  & 1.18  & 10.8  & 2700  &   24 & 7.43$\pm$0.09 & 7.90$\pm$0.10 & 0.07 &  0.14$\pm$0.13 \\
 2597 &  51330 &               & F2 Ib-II &  6710  & 2.02  &  3.3  &  935  &  7.1 & 7.32$\pm$0.13 & 8.15$\pm$0.29 & 0.32 &  0.50$\pm$0.32 \\
 2839 &  58585 &               & A8 I-II  &  7240  & 1.92  &  2.0  & 1330  &  8.6 & 7.40$\pm$0.14 & 8.39$\pm$0.14 & 0.56 &  0.66$\pm$0.20 \\
 2874 &  59612 &               &  A5 Ib   &  8620  & 1.78  &  7.8  &  940  & 12.9 & 7.52$\pm$0.18 & 8.24$\pm$0.10 & 0.41 &  0.39$\pm$0.21 \\
 2933 &  61227 &               &  F0 II   &  6690  & 2.02  &  2.7  &  790  &  7.0 & 7.37$\pm$0.13 & 8.40$\pm$0.17 & 0.57 &  0.70$\pm$0.21 \\
 3291 &  70761 &               &  F3 Ib   &  6600  & 1.25  &  3.9  & 2900  & 14.2 & 7.41$\pm$0.16 & 8.65$\pm$0.15 & 0.82 &  0.91$\pm$0.22 \\
 6144 & 148743 &               &  A7 Ib   &  7400  & 1.80  &  4.8  & 1330  & 10.0 & 7.39$\pm$0.16 & 8.60$\pm$0.23 & 0.77 &  0.88$\pm$0.28 \\
 7014 & 172594 &               &  F2 Ib   &  6760  & 1.66  &  4.6  & 1000  & 10.0 & 7.43$\pm$0.15 & 8.36$\pm$0.16 & 0.53 &  0.60$\pm$0.22 \\
 7094 & 174464 &               &  F2 Ib   &  6730  & 1.75  &  3.4  &  855  &  9.1 & 7.31$\pm$0.16 & 8.15$\pm$0.16 & 0.32 &  0.51$\pm$0.23 \\
 7387 & 182835 &               &  F3 Ib   &  6700  & 1.43  &  4.4  &  880  & 12.5 & 7.47$\pm$0.11 & 8.61$\pm$0.13 & 0.78 &  0.81$\pm$0.17 \\
 7770 & 193370 &               &  F5 Ib   &  6180  & 1.53  &  5.0  &  960  & 10.0 & 7.28$\pm$0.07 & 8.14$\pm$0.12 & 0.31 &  0.53$\pm$0.14 \\
 7823 & 194951 &               &  F1 II   &  6760  & 1.92  &  4.2  & 1010  &  7.8 & 7.37$\pm$0.12 & 8.43$\pm$0.14 & 0.60 &  0.73$\pm$0.18 \\
 7847 & 195593 &               &  F5 Iab  &  6290  & 1.44  &  4.1  & 1040  & 11.2 & 7.44$\pm$0.15 & 8.34$\pm$0.14 & 0.51 &  0.57$\pm$0.21 \\
 7876 & 196379 &               &  A9 II   &  7020  & 1.66  &  3.4  & 1740  & 10.6 & 7.29$\pm$0.16 & 7.75$\pm$0.13 & -0.08 &  0.13$\pm$0.21 \\
\hline
\end{tabular}
\end{minipage}
\end{table*}
      
      We found that the non-LTE corrections to the N abundance are negative for all used N I lines. They depend on the effective temperature $T_{\rm eff}$ , surface gravity $\log g$ and microturbulent parameter $V_t$ , as well as on the N abundance itself. There is an evident trend with $T_{\rm eff}$, so the higher $T_{\rm eff}$ the greater the non-LTE corrections. Moreover, the non-LTE corrections tend to be greater for the stronger N I lines. We show below some non-LTE effects on the nitrogen abundances.

\section{Determination of the nitrogen abundances}

      Following Paper I we divided 30 programme A- and F-type supergiants into two groups according to their distances $d$. As mentioned above, for stars with  $d < 700$ pc an accuracy of parameters $T_{\rm eff}$ and $\log g$ is especially high. It should be noted that, as shown in Paper I, the iron abundance of the supergiants, i.e. their metallicity coincides virtually with the solar one.

      We present in Table 3 for 16 near supergiants and 14 distant ones the following data: HR an HD numbers of stars, their names (if they exist) and spectral classification. Next we provide the stellar parameters from Paper I, namely the effective temperature $T_{\rm eff}$, surface gravity $\log g$, microturbulent parameter $V_t$, distance $d$, mass $M$ in solar masses $M_\odot$ and the iron abundance $\log\epsilon$(Fe). It should be noted that the latter value, as well as the nitrogen abundance $\log\epsilon$(N) is given in the usual logarithmic scale, where for hydrogen the value $\log\epsilon$(H) = 12.00 is adopted.
      
      Model atmospheres for the stars were computed using Kurucz's (1993) code ATLAS9 for the adopted $T_{\rm eff}$, $\log g$ and $V_t$ values. The normal helium abundance He/H = 0.10, as well as the normal (star) metallicity was adopted in the computations (Note that the He/H evaluation for programme stars is impossible due to their low temperatures, $T_{\rm eff} \leq 8600$~K). Using these model atmospheres and the microturbulent parameter $V_t$ from Table 3, we determined the nitrogen abundance $\log\epsilon$(N) for each star. When $\log\epsilon$(N) deriving, we used the measured equivalent widths $W$ presented in Table A1 for 22 programme stars; the $\log\epsilon$(N) values for the individual N I lines are shown in Table A1, too. Note that these abundances were always checked with the help of synthetic spectra computations (i.e. from line profiles). The computed profiles of the N I lines were the principal source of $\log\epsilon$(N) for 13 remaining stars with relatively high rotational velocities or with blended spectra, when accurate $W$ measurements were problematic. As mentioned above, the synthetic spectra were also applied when the NI lines 8680.28, 8683.40 and 8686.14 \AA\ in the wing of the $P_{13}$ line were analyzed.
      
      The derived mean abundances $\log\epsilon$(N) for all 30 stars are presented in Table~3. The tabulated errors in $\log\epsilon$(N)  incorporate four contributors, namely the scatter in the abundance estimates from the various N I lines and the uncertainties in the parameters $T_{\rm eff}$ , $\log g$ and $V_t$. In the case of F supergiants, all four contributors need to be considered. In the case of A supergiants, the main sources of the errors are the scatter between lines and the uncertainties in $T_{\rm eff}$ , with the errors in $\log g$ and $V_t$ playing a minor role. Only for some distant supergiants, when the uncertainty in $\log g$ may exceed $\pm$0.2 dex, is the contribution from $\log g$ a factor. It should be noted that there is no significant difference in accuracy of the $\log\epsilon$(N) values for near and distant supergiants. In particular, the mean error is $\pm$0.12 dex for the first group and $\pm$0.15 dex for the second one.
      
      We provide in the final columns of Table 3 two quantities, which give the N abundances relative to the Sun, namely: $[N/H]$ = $\log\epsilon$(N) - $\log\epsilon_\odot$(N) and $[N/Fe]$ = $\log\epsilon$(N/Fe) - $\log\epsilon_\odot$(N/Fe), where $\log\epsilon$(N/Fe) = $\log\epsilon$(N) - $\log\epsilon$(Fe). The value $[N/Fe]$ takes into account the difference in metallicities of the stars. We used the solar abundances from Asplund et al. (2009), specifically $\log\epsilon_\odot$(N) = 7.83$\pm$0.05 and $\log\epsilon_\odot$(Fe) = 7.50$\pm$0.04; therefore, $\log\epsilon_\odot$(N/Fe) = 0.33. Recently Caffau et al. (2009) obtained the solar nitrogen abundance $\log\epsilon_\odot$(N) = 7.86$\pm$0.12 that is very close to Asplund et al.'s value. Note that these estimates of the solar abundances are based on the three-dimensional (3D) hydrodynamical model of the solar atmosphere. It is interesting that our estimation of the solar N abundance determined with the above-described technique and the 1D-model of the solar atmosphere (Castelli \& Kurucz 2003) provided $\log\epsilon_\odot$(N) = 7.89$\pm$0.05 that is greater only by 0.06 dex than Asplund et al.'s value. Note that all these small differences in $\log\epsilon_\odot$(N) cannot play any marked role for the interpretation of the derived significant N overabundances in the supergiants, which attain 0.8-0.9 dex. 
      
      We do not indicate in Table 3 the uncertainties in the $[N/H]$ values, because they are the same as in $\log\epsilon$(N). As far as the uncertainties in the $[N/Fe]$ values are concerned, they should depend on errors in both $\log\epsilon$(N) and $\log\epsilon$(Fe), i.e. $\sigma$(N) and $\sigma$(Fe). We present in Table 3 the resultant errors $\sigma([N/Fe]) = \sqrt{\sigma(N)^2 + \sigma(Fe)^2}$. However, it should be noted that these values can be greater than the real errors in $[N/Fe]$. In fact, when the value $[N/Fe]$ is determined, the errors in $\log\epsilon$(N) and $\log\epsilon$(Fe) due to the uncertainties in $\log g$ reveal a mutual (at least, partial) elimination, because both $\log\epsilon$(N) and $\log\epsilon$(Fe) show the similar dependence on $\log g$. This is the case too for uncertainties in $V_t$. On the other hand, this is not the case for the errors in $\log\epsilon$(N) and $\log\epsilon$(Fe) due to the uncertainties in $T_{\rm eff}$ , because the dependence of $\log\epsilon$(N) and $\log\epsilon$(Fe) on $T_{\rm eff}$ is different. (The N and Fe abundances are derived from N I and Fe II lines, respectively). 

      It is interesting to compare the derived non-LTE nitrogen abundances with the LTE ones. In Fig.4, we show the difference (LTE - non-LTE) as a function of $T_{\rm eff}$ for all programme stars. Evidently, there is a trend with $T_{\rm eff}$: the higher $T_{\rm eff}$ the greater the difference on average. The difference is about 0.2-0.3 dex for supergiants with $T_{\rm eff}$ = 5800-6400 K, 0.3-0.6 dex for stars with $T_{\rm eff}$ = 6700-7400 K and 0.6-0.9 dex for stars with $T_{\rm eff}$ around 8500 K. The scatter for stars with similar effective temperatures $T_{\rm eff}$ can be explained by either a marked difference in the N abundances (stars with $T_{\rm eff}$ between 6500 and 7100 K) or a significant difference in the microturbulent parameter $V_t$ (A-type stars with $T_{\rm eff}$ around 8500 K). Note that the higher $\log\epsilon$(N) or $V_t$ the greater the difference (LTE - non-LTE). 

\begin{figure}
\epsfxsize=8truecm
\epsffile{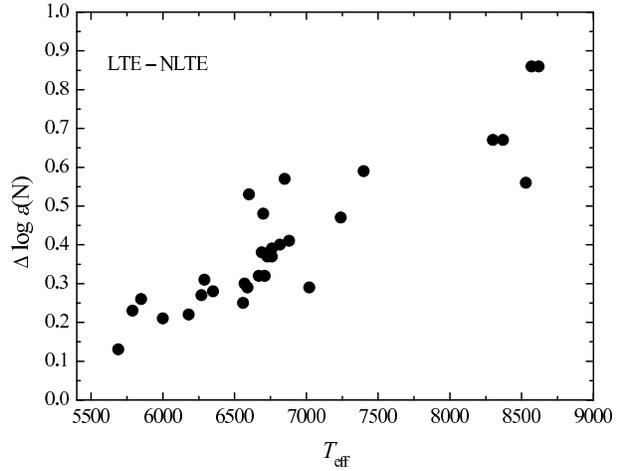}
\caption{The difference between the LTE and non-LTE mean nitrogen abundances for the 30 supergiants as a function of their effective temperature.}
\end{figure}

\begin{figure}
\epsfxsize=8truecm
\epsffile{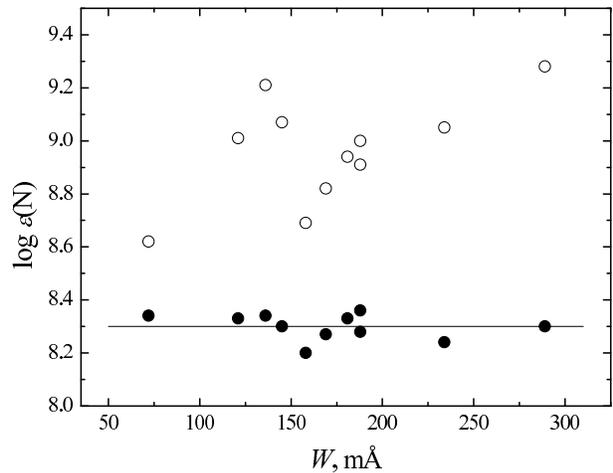}
\caption{The N abundances determined for the A-type supergiants HR~6081 as a function of equivalent widths of N~I lines. LTE abundances - unfilled circles, non-LTE ones - filled circles. The thin straight line corresponds to the mean non-LTE abundance 8.30.}
\end{figure}

      In Fig.5 we show for the A-type star HR 6081 with $T_{\rm eff}$ = 8370 K both the LTE and non-LTE abundances derived from individual N I lines as a function of the equivalent width W. The LTE abundances display a large scatter and a trend with W. In sharp contrast, the non-LTE abundances lie very compactly around the mean value $\log\epsilon$(N) = 8.30, and the trend with $W$ is absent. One sees that the non-LTE corrections for the rather strong N I lines can reach 0.9-1.0 dex. Note that the absence of the trend with $W$ shows that the microturbulent parameter $V_t$ = 2.8 km s$^{-1}$ determined for HR 6081 in Paper I from Fe II lines is a suitable value as well for N I lines. Figs. 4 and 5 confirm that the non-LTE approach is necessary for the N I line analysis and that the adopted approach is likely yielding the true N abundances. 

\begin{figure}
\epsfxsize=8truecm
\epsffile{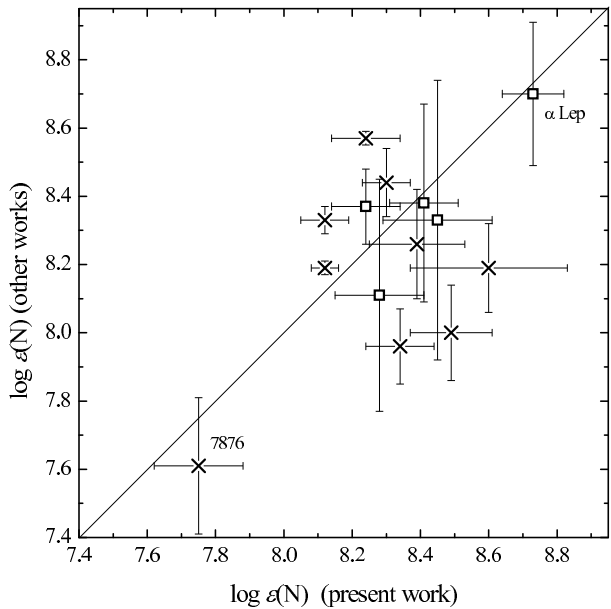}
\caption{Comparison of our non-LTE nitrogen abundances with the non-LTE abundances from TT'95 (squares) and VP'03 (crosses).}
\end{figure}

      In Fig.6 we compare our $\log\epsilon$(N) values with the results of TT'95 and VP'03 for those stars in common, all are non-LTE values. One sees that there is rather good agreement with the N abundances of TT'95 for 5 common stars (squares). Virtually the same great $\log\epsilon$(N) value is found by us and TT'95 for $\alpha$Lep (marked in Fig.6). In the case of VP'03 a marked scatter is seen; discrepancies attain 0.4-0.5 dex (crosses). We excluded from the comparison with VP'03 one of 10 common stars, HD 36673 ($\alpha$Lep), because only one N I line was used by VP'03; the N abundance for $\alpha$Lep was significantly underestimated by VP'03 in comparison with us and TT'95. Of the nine remaining common stars, the distant supergiant HR 7876 is of special interest (marked in Fig.6). Both works give a low nitrogen abundance, namely $\log\epsilon$(N) = 7.75$\pm$0.13 by us and $\log\epsilon$(N) = 7.61$\pm$0.20 by VP'03, values that are close to the solar abundance $\log\epsilon_\odot$(N) = 7.83.
As mentioned above, the discrepancy up to 0.4-0.5 dex between our and VP'03 results is mostly explained by the difference in parameters $T_{\rm eff}$, $\log g$ and $V_t$.

\begin{figure}
\epsfxsize=8truecm
\epsffile{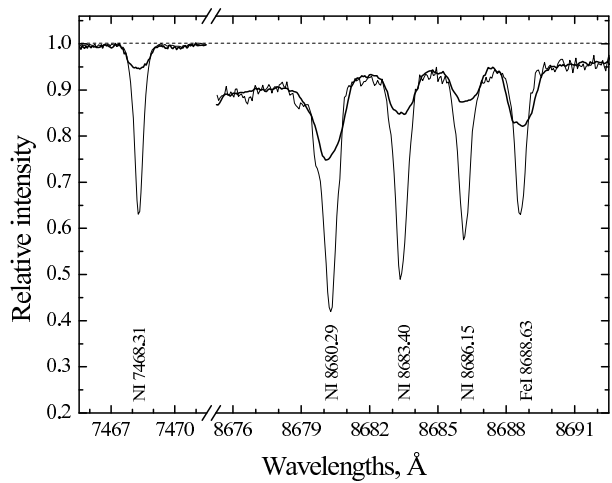}
\caption{Comparison of N~I spectra for two stars with different nitrogen abundances, namely HR~7876 (thick line) and $\alpha$~Lep = HR~1865 (thin line). The N~I isolated line at 7468.31 \AA, as well as three N~I lines in the $P_{13}$ wing are shown.}
\end{figure}

      The large difference in the N abundance between $\alpha$Lep and HR~7876 is evident from their spectra. We show in Fig.7 for both stars the isolated line N I 7468.31 \AA, as well as the spectral region with three N I lines at 8680.29, 8683.40 and 8686.15 \AA\ in the $P_{13}$ wing. The parameters $T_{\rm eff}$ and $\log g$ for the stars are rather close (Table 3), but the N I lines for HR 7876 are much weaker than for $\alpha$Lep. There is some difference in rotational velocities, namely $v$sin$i =$ 10 km s$^{-1}$ for $\alpha$Lep and 25 km s$^{-1}$ for HR~7876 (our preliminary estimates); however, inspite of this, the conclusion about the weakness of N I lines in the spectrum of HR~7876 remains in force. Note that the line Fe I 8688.63 \AA\ is shown as well in Fig.7 but with a similar $W$ for these stars, namely $W$ = 217 and 185 m\AA\ for $\alpha$Lep and HR 7876, respectively; this difference in Ws is fully explained by small differences in $T_{\rm eff}$ and $\log\epsilon$(Fe) (Table 3).     

\section{Nitrogen abundance as a function of parameters $T\lowercase{_{\rm eff}}$, $\log\lowercase{{g}}$ and $M/M_\odot$} 

      An observer's initial search for an understanding of the N overabundances naturally looks for correlations with basic observable stellar parameters: effective temperature $T_{\rm eff}$ and surface gravity $\log g$. In searching for trends with $T_{\rm eff}$ and $\log g$, we considered both our results and data of TT'95 and VP'03. We included 20 stars from VP'03 (two stars for which a single N I line was used are omitted) and 10 stars with $T_{\rm eff} <$ 10000 K from TT'95. Nitrogen abundances $[N/H]$ for the entire dataset are shown in Fig.~8 against $T_{\rm eff}$ and in Fig.~9 against $\log g$. In both figures, the dashed line at $[N/H]$~=~0 corresponds to the solar abundance of 7.83. Our data for near and distant supergiants are presented by filled and open circles, respectively.
Note that typical errors in the N abundances are $\pm$0.12 and $\pm$0.15 for near and distant stars, respectively, whereas the averaged errors are $\pm$0.08 for 20 stars from VP'03 and $\pm$0.27 for 10 stars from TT'95 (two latter values are found from the error estimations presented for individual stars in VP'03 and TT'95).

      One sees that across the $T_{\rm eff}$ interval from 5500K to 10,000 K and the $\log g$ interval from 0.6 to 2.6 dex, nitrogen is overabundant except for the two stars, namely HR 825 and HR 7876. No trends are discernible, but the degree of overabundance covers a large spread, say 0.2 to 0.9 dex.

\begin{figure}
\epsfxsize=8truecm
\epsffile{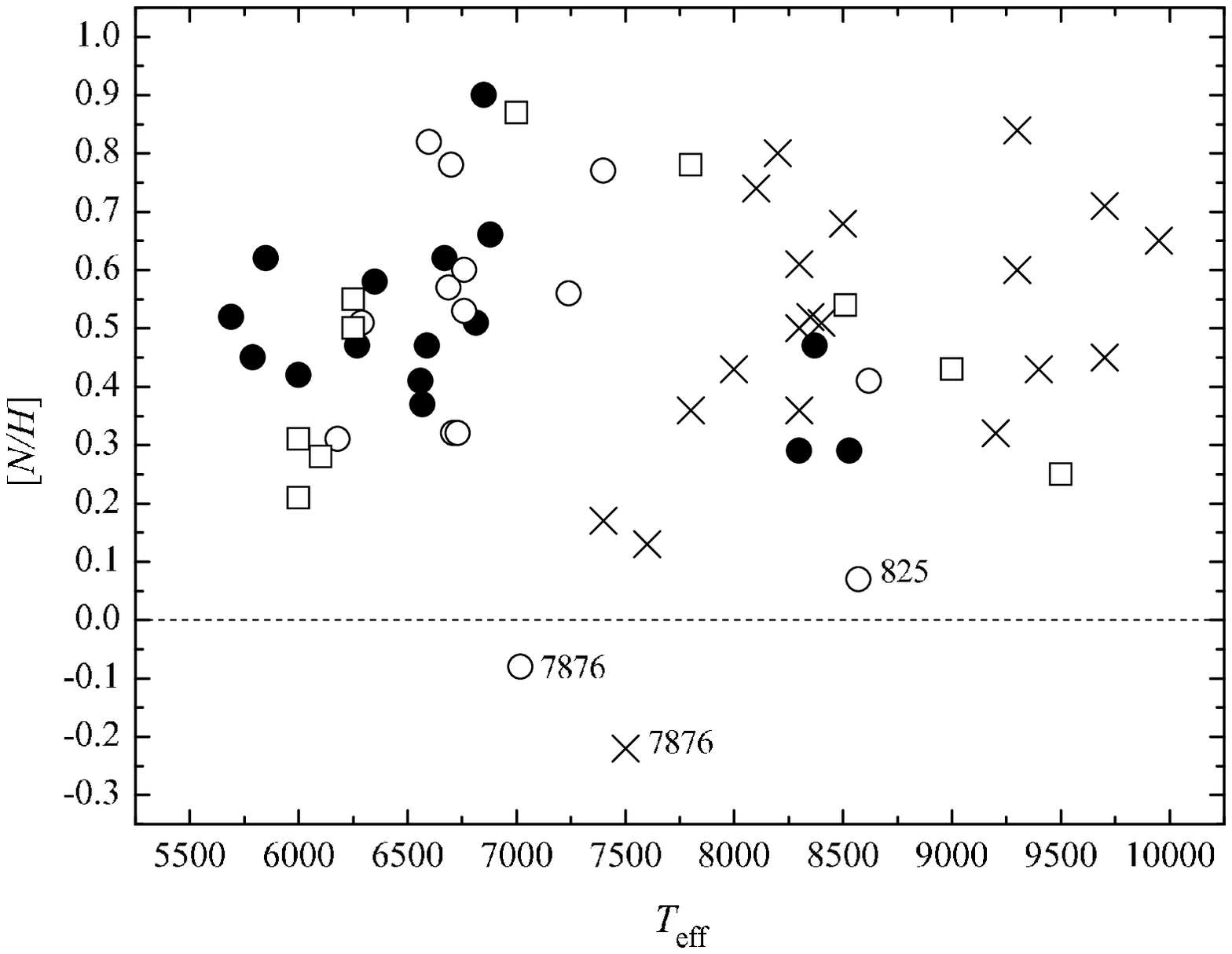}
\caption{The nitrogen excess $[N/H]$ as a function of the effective temperature. Our data for near supergiants with $d < 700$ pc and for distant supergiants with $d > 700$ are presented by filled and open circles, respectively. Squares and crosses correspond to data of TT'95 and VP'03, respectively. Dashed zero line corresponds to the Sun.}

\epsfxsize=8truecm
\epsffile{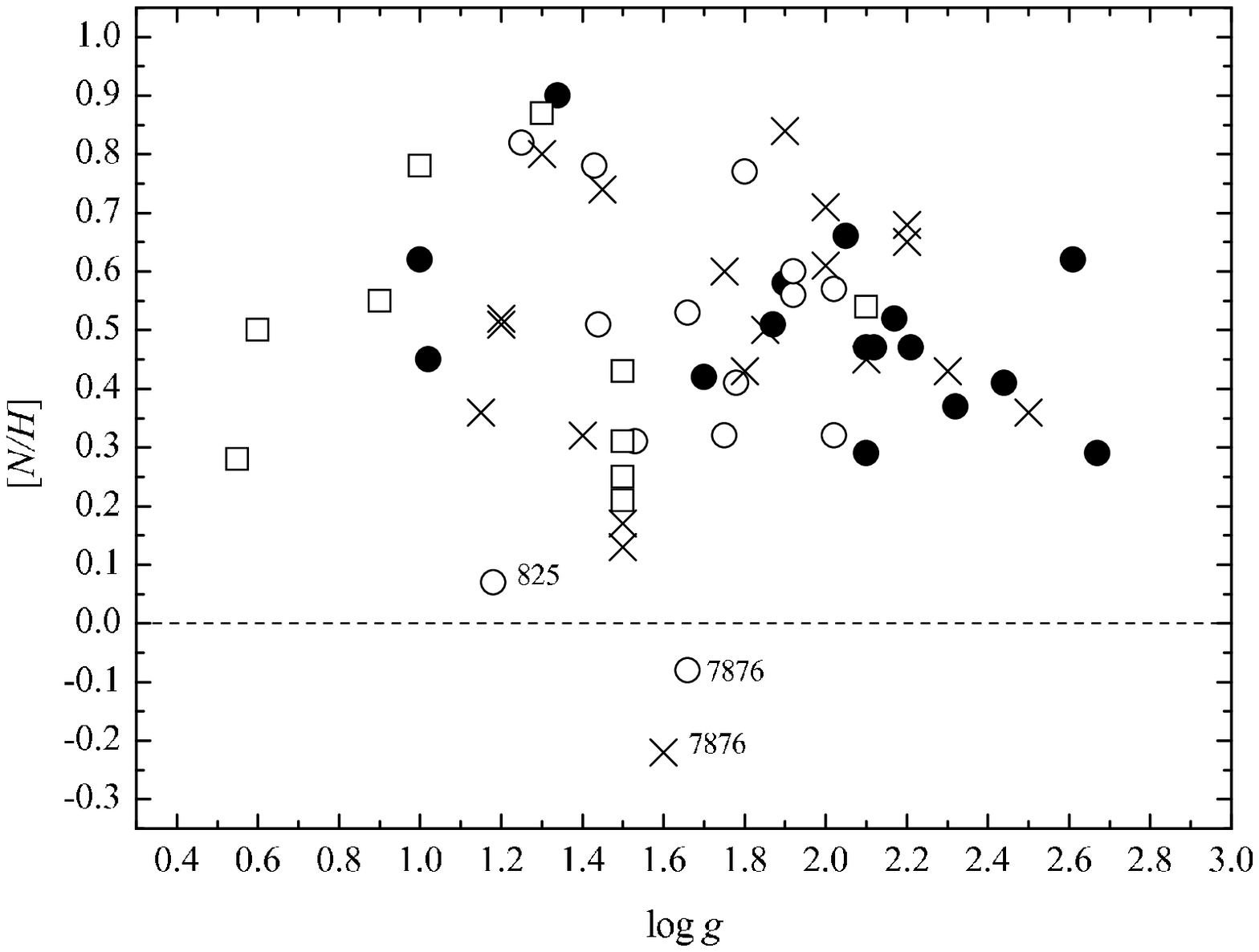}
\caption{The nitrogen excess $[N/H]$ as a function of the surface gravity. Symbols are the same as in Fig.8.}
\end{figure}

      Stellar mass $M$ is obviously a parameter of great interest, as evolution of stars depends strongly on $M$. We considered at first a relation between $[N/H]$ and $M/M_\odot$ for supergiants from TT'95 and VP'03. It should be noted that we determined the masses $M/M_\odot$ of these stars, like $M/M_\odot$ of our programme stars (Paper I), from Claret's (2004) evolutionary tracks, basing on the star's parameters $T_{\rm eff}$ and $\log g$ from TT'95 and VP'03. We found that the $M/M_\odot$ values vary between 9.1 and 25.5 for 10 stars from TT'95 and between 6.0 and 21.8 for 20 stars from VP'03, so the $M/M_\odot$ spread is rather wide. Nevertheless, no discernible trend of $[N/H]$ with $M/M_\odot$ is found.
We suppose that errors in $T_{\rm eff}$ and $\log g$ can play a significant role here. In particular, the $\log g$ values tend to be markedly underestimated by TT'95. For instance, too low $\log g$ were adopted by TT'95 for the bright F-type supergiants $\alpha$~Per, $\delta$~CMa and $\gamma$~Cyg, as compared with our accurate $\log g$ values derived from stellar parallaxes (note that the parallaxes for these stars are obtained with errors $\pm$3, $\pm$19 and $\pm$15 per cent, respectively, see Paper I).
This led to the small lowering of the N abundances but, at the same time, to the strong $M/M_\odot$ overestimation ($M/M_\odot$ = 17.3, 25.1 and 25.5, whereas our values are 7.3, 14.9 and 14.5 for $\alpha$Per, $\delta$CMa and $\gamma$Cyg, respectively). Such errors in $M/M_\odot$ could mask a possible trend of $[N/H]$ with $M/M_\odot$.

\begin{figure}
\epsfxsize=8truecm
\epsffile{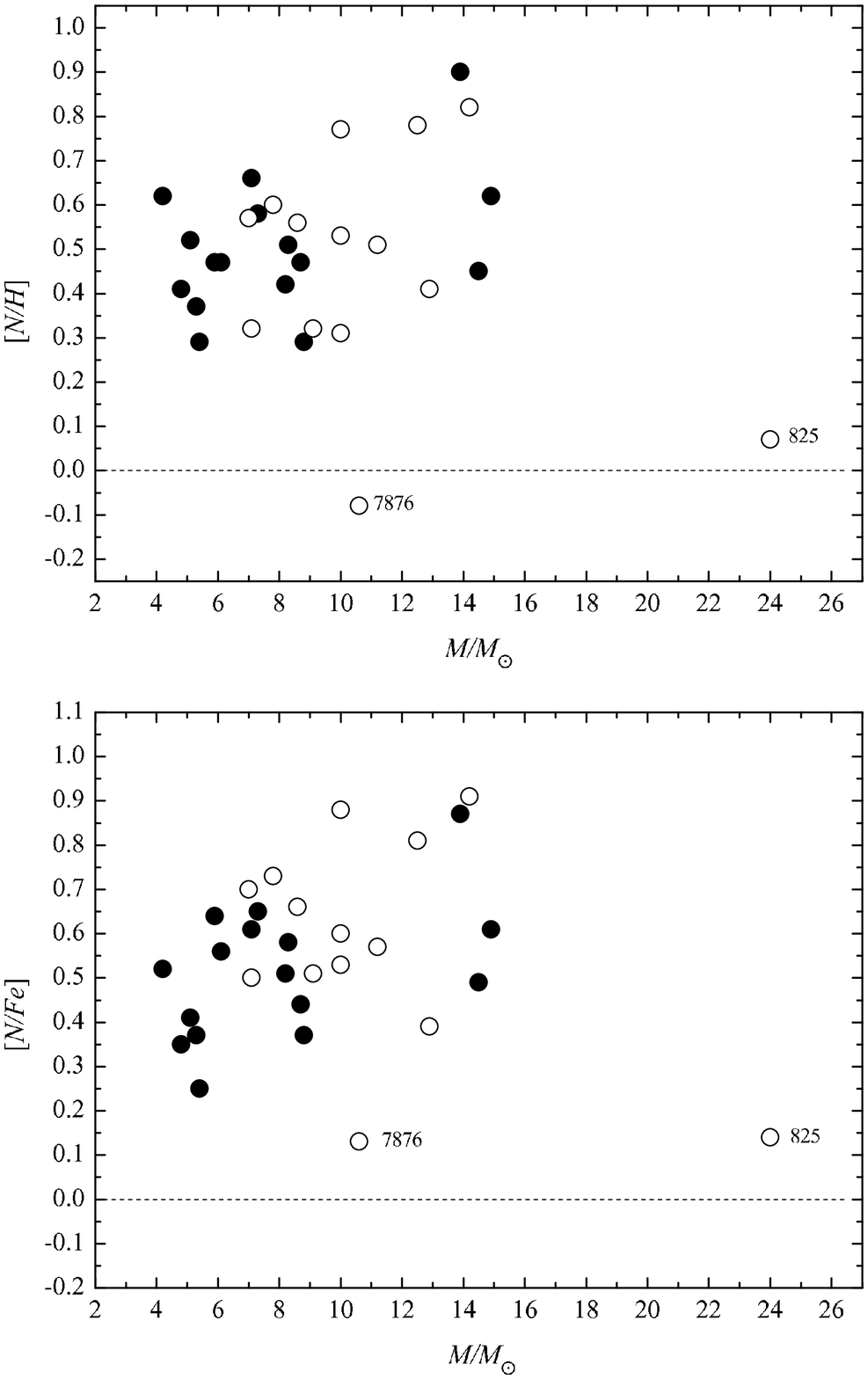}
\caption{Our values $[N/H]$ and $[N/Fe]$ as a function of the mass. Near and distant supergiants are shown by filled and open circles, respectively. Dashed zero lines correspond to the Sun.}
\end{figure}

      Fig.10 shows our $[N/H]$ and $[N/Fe]$ values as a function of the mass $M/M_\odot$. One may see that, if obvious 'outliers' HR 825 and HR 7876 are excluded, all supergiants occupy a rather compact region with the masses between 4 and 15 $M_\odot$ and the N excesses from 0.2-0.3 to 0.9 dex. One may see that there is no systematic discrepancy between the near and distant supergiants (filled and open circles, respectively), so we shall not make a difference between these two groups in next figures. It is important that the $[N/H]$ and $[N/Fe]$ values for the low-mass supergiants are less on average than for the more massive stars. In particular, the $[N/Fe]$ values for 4-5 $M_\odot$ stars vary from 0.2 to 0.5 dex, whereas for 12-15 $M_\odot$ stars they vary from 0.4 to 0.9 dex. So, there is a weak apparent trend in the nitrogen excess due to such a difference. 
      
      The 'outliers', HR~825 and HR~7876, are of special interest. Parallaxes of these distant supergiants could not be measured with confidence and, therefore, could not be used for the $\log g$ determination. Nevertheless, parameters $T_{\rm eff}$ and $\log g$ of these stars have been derived in Paper~I with rather good accuracy, namely: $T_{\rm eff}$~=~8570$\pm$160 K; $\log g$~=~1.18$\pm$0.13 for HR~825 and $T_{\rm eff}$~=~7020$\pm$100 K; $\log g$~=~1.66$\pm$0.10 for HR~7876. There are few earlier determinations of these parameters for the stars. Verdugo et al. (1999) found for HR 825 the effective temperature $T_{\rm eff}$~=~8510 K that is very close to our value. The same temperature $T_{\rm eff}$~=~8510 K was adopted for HR~825 by Lamers et al. (1995); from their estimations of the mass and  radius, $M/M_\odot$ = 20 and $R/R_\odot$~=~173, we obtained $\log g$~=~1.26 that is in very good agreement with our value $\log g$~=~1.18. For the supergiant HR~7876 two previous determinations of $T_{\rm eff}$ and $\log g$ are known to us, namely $T_{\rm eff}$~=~7190 K; $\log g$~=~2.09 (Gray et al. 2001) and $T_{\rm eff}$~=~7500 K; $\log g$~=~1.6 (Venn 1995). One may note a rather good agreement with our $T_{\rm eff}$ in the first work and with our $\log g$ in the second one. 
      
      The accuracy of the derived nitrogen abundances for HR~825 and HR~7876 is rather high, too, namely $\log\epsilon$(N)~=~7.90$\pm$0.10 for HR~825 and $\log\epsilon$(N)~=~7.75$\pm$0.13 for HR~7876. As mentioned above, in the case of HR~7876 its low N abundance was confirmed by VP'03. Summarizing, we may conclude that the supergiants HR~825 and HR~7876 have really an unusually low nitrogen abundance among the supergiants, which is close to the solar value 7.83$\pm$0.05. The slight overabundances $[N/Fe]$~=~0.14$\pm$0.13 and 0.13$\pm$0.21, respectively (Table 3) are within the uncertainties. So, one may suppose that these two 'outliers' have virtually their initial N abundance. 

\begin{figure}
\epsfxsize=8truecm
\epsffile{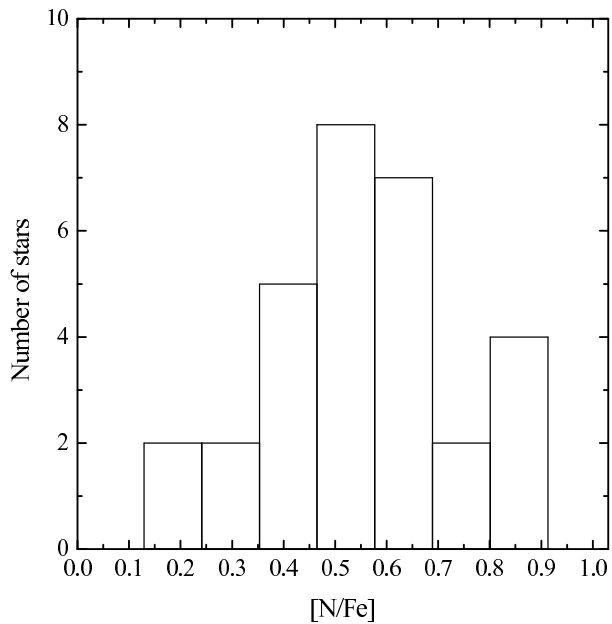}
\caption{Distribution of programme stars on the $[N/Fe]$ values.}
\end{figure}

      One may see from Fig.10 that the distribution of 30 programme supergiants on $[N/Fe]$ is uneven. In fact, most of the stars display the N excess between 0.4 and 0.7 dex. The uneven distribution is visible from a histogram presented in Fig.11. One may conclude that the nitrogen enrichment of the supergiant atmospheres occurs preferably with $[N/Fe]$~=~0.4-0.7.

\section{Comparison of the observed and predicted nitrogen abundances}

      An enhancement of the N abundance is evidently a general property of the A- and F-type supergiants. The N overabundance in A, F and G supergiants was anticipated from previous studies. Note that the N excess is measured with respect to the solar N abundance. Strictly speaking, one should reference the N enhancement to the N abundance in early B-type main sequence (MS) stars, the immediate progenitors of these supergiants. It is an empirical result that the B-type MS stars and the Sun have very similar chemical compositions (see below). Therefore, the N enhancement of supergiants may really be measured with respect to the solar abundance. 

\begin{figure}
\epsfxsize=8truecm
\epsffile{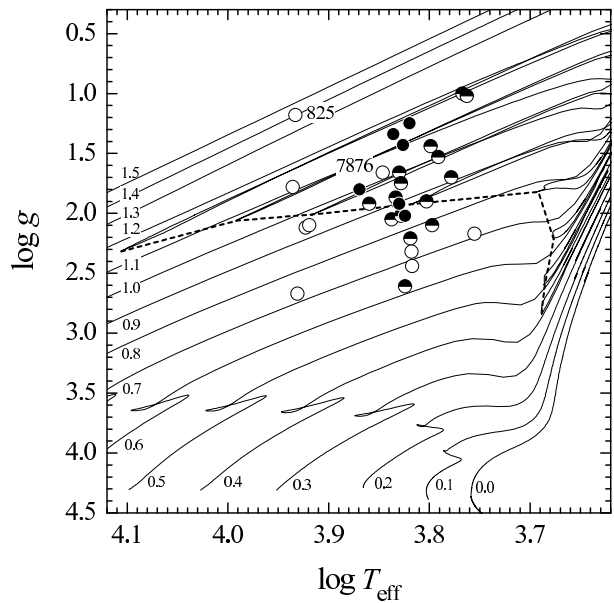}
\caption{The $\log T_{\rm eff}$ - $\log g$ diagram. Programme stars are plotted according to their ($T_{\rm eff}$, $\log g$) from Table 3. Filled circles correspond to the stars with $[N/Fe]$ = 0.70-0.91, half-filled circles - $[N/Fe]$ = 0.49-0.66 and open circles - $[N/Fe]$ = 0.13-0.44. Two stars without the discernible N excess, HR~825 and HR~7876, are marked. Evolutionary tracks from Claret (2004) are shown for values $\log M/M_\odot$ from 0.0 to 1.5. Dashed line connects the ends of the red-blue loops.}
\end{figure}

      In Fig.12 we show 30 programme supergiants in the log $T_{\rm eff}$ - $\log g$ diagram, on which are superimposed evolutionary tracks from Claret (2004); the tracks were computed for the initial composition X = 0.70 and Z = 0.02. The supergiants are divided into three groups according to their $[N/Fe]$ values, namely $[N/Fe]$ = 0.70-0.91 (filled circles, 6 stars), $[N/Fe]$ = 0.49-0.66 (half-filled circles, 15 stars) and $[N/Fe]$ = 0.13-0.44 (open circles, 9 stars). Note that the middle group is most numerous, that is in accordance with the histogram in Fig.11. One may suppose that some supergiants on the $T_{\rm eff}$ - $\log g$ diagram can be objects, which terminated recently the MS phase and evolve now to the red giant/supergiant phase (i.e. they are post-MS objects). Other supergiants could already pass the latter phase; now they move along the red-blue loops on evolutionary tracks, at first blueward and then again redward. It would be interesting to separate these two kinds of supergiants, because their surface N enrichment can be of different origin. 

      Three processes have been suggested to result in the surface N enrichment for massive stars. In each case, the enrichment results from contamination of the atmosphere with CN-cycled material (i.e. matter rich in N and poor in C) from the interior. The proposed processes invoked in the case of single stars are the severe mass loss and the rotationally induced mixing during the MS phase, as well as the dredge-up associated with the deep convective envelope of a cool supergiant. Mass loss is believed to be unimportant for the mass range covered by the B-type stars that are the progenitors of the supergiants discussed here (masses of our programme supergiants are $M$ = 4-15 $M_\odot$, except for HR 825). Effects of two remaining processes are somewhat coupled; for example, the consequences of the first dredge-up (FD) in a rotating and a non-rotating star of otherwise identical properties will be different. 

      With the exception of the FD, the processes may begin to make their mark in the MS stars. Alternatively, if the FD is the sole effective process altering the surface N abundance, alterations to the atmosphere's N abundance first appear among G-K supergiants. The N enrichment and C depletion result from CN-cycled material being mixed to the surface by the convective envelope. Following the FD, stars may evolve along the red-blue loops with an atmosphere having the abundance changes that were brought about by the FD. A star on the loop evolves more slowly than the post-MS star of the same mass. Thus, in the region of the $T_{\rm eff}$ - $\log g$ diagram mapped out by the red-blue loops, the majority of supergiants should be really on the loops and, therefore, their N enrichment is (entirely or partially) a result of deep mixing during the FD. 

      It is obviously that the post-FD objects can be located only inside the red-blue loop area. One may see from Fig.12 that, according to Claret's (2004) computations, extensive red-blue loops occur for stars with masses from 6 to 13 $M_\odot$. Dashed line in Fig.12 connects the blue ends of the loops; so, this line is a lower boundary of the loop area. El Eid (1995) considered carefully the problem of the loops and showed that properties of the loops (their location and extension) depend on many factors. When using El Eid's evolutionary tracks, we found that the lower boundary, i.e. dashed line in Fig.12 should be shifted downwards by about 0.4 dex. The much greater uncertainty takes place for the upper boundary of the loop area. In fact, according to El Eid, an existence of the loops for relatively massive stars depends on the criterion used in computations of convective mixing, namely: the Schwarzschild criterion gives the loops up to $M$ = 13 $M_\odot$, whereas the Ledoux criterion gives the loops up to $M$ = 19 $M_\odot$. So, the loop area can be significantly expanded as compared with Fig.12. This supposition is in agreement with the recent Przybilla et al.'s (2009) conclusion based on a study of A and B supergiants with temperatures $T_{\rm eff}$ = 8000-15000 K, namely: 'Blue loops are suggested to extend to higher masses and to higher $T_{\rm eff}$ than predicted by the current generation of stellar evolution models'. Apropos, Kippenhahn \& Weigert (1990) wrote twenty years ago that the red-blue loop phase is a sort of magnifying glass, revealing relentlessly the faults of calculations of earlier phases'.  

\begin{figure}
\epsfxsize=8truecm
\epsffile{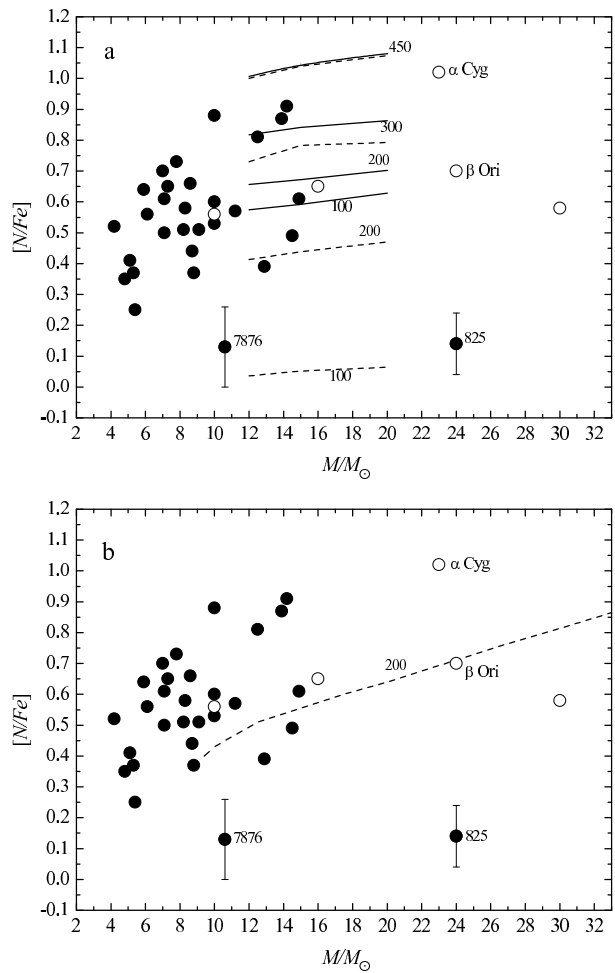}
\caption{Comparison of the $[N/Fe]$ vs. $M/M_\odot$ relation with theoretical predictions of (a) Heger \& Langer (2000) and (b) Maeder et al. (2009). Dashed lines correspond to the post-MS phase, solid lines - the post-FD phase. For case (a) four initial rotational velocities are considered, approximately 100, 200, 300 and 450 km s$^{-1}$; for case (b) only one post-MS phase line is given for an initial rotational velocity of about 200 km s$^{-1}$. Filled circles show our data, open circles display data of Przybilla et al. (2006) and Schiller \& Przybilla (2008) for five A and B supergiants.}
\end{figure}

      Summarizing, we may conclude that using the $\log T_{\rm eff} - \log g$ diagram presented in Fig.12 we cannot separate the post-MS and post-FD supergiants, because the boundaries of the loop area are uncertain. Now we seek to relate the surface N enrichments of the supergiants to predictions for rotationally-induced mixing in B-type MS stars, the progenitors of the A- and F-type supergiants. Heger \& Langer (2000) computed stellar models with the rotationally induced mixing during the MS phase and evaluated changes in the surface abundances of several light elements, including nitrogen, both at the MS termination and after the FD. We present in Fig.13a their results for models with masses $M$ = 12, 15 and 20 $M_\odot$. Four values of the initial rotational velocity $v_0$ are considered, which are approximately 100, 200, 300 and 450 km s$^{-1}$. Two curves are shown for each $v_0$ value, namely: the dashed line corresponds to the post-MS phase and the solid one corresponds to the post-FD phase. Our $[N/Fe]$ values as a function of $M/M_\odot$ are shown there as well (filled circles). The $[N/Fe]$, as noted above, are less sensitive to uncertainties affecting the stellar parameters than $[N/H]$. Moreover, data of Przybilla et al. (2006) and Schiller \& Przybilla (2008) for five A and B supergiants are presented, too (open circles). It should be noted that these $[N/Fe]$ values, like ours, are based on the N and Fe abundances derived from N I and Fe II lines, respectively. One sees that most of our stars are less massive than the lowest mass (12 $M_\odot$) considered by Heger \& Langer, and three of five stars from two works cited, as well as HR~825 are more massive than their highest mass (20 $M_\odot$). However, in light of the weak mass dependence of the predictions, we consider an informative comparison of observation and prediction is possible. 

      It is important to note that the $v_0$ value is the velocity at the beginning of the MS phase. When a star terminates this evolutionary phase and passes into the AFG supergiant phase, its rotational velocity decreases significantly. In particular, according to our preliminary estimates, all 30 supergiants from our list have the projected rotational velocities $v$sin$i <$ 60 km s$^{-1}$ and for most of them $v$sin$i <$ 20 km s$^{-1}$. 

      One may see from Fig.13a that a difference between the two cases, namely post-MS and post-FD, depends strongly on the initial velocity $v_0$. In particular, for $v_0$ = 100 km s$^{-1}$ the surface N abundance at the post-MS phase (dashed line) is not markedly enriched but N is enriched at the post-FD phase (solid line). One implication is that the rotationally induced mixing of CN-cycled material is either unimportant for at velocities of less than about 100 km s$^{-1}$ and/or becomes only observable after the FD phase. At the other extreme,  for rotational velocities of about 300 km s$^{-1}$ and higher, the full effect of rotationally induced mixing on the surface N abundance is apparent in the post-MS phase with the FD phase making little or no extra enhancement of the N abundance.
      
      In Fig.13b we present theoretical results from another source; namely, we reproduce there a part of the curve from Maeder et al. (2009, see their Fig.1), where the post-MS nitrogen excess in atmospheres of 10-33 $M_\odot$ stars with the approximate initial velocity $v_0$ = 200 km s$^{-1}$ is shown. When comparing Figs. 13a and 13b, one may see a rather good agreement between both dashed lines for $v_0$ = 200 km s$^{-1}$.

      Apropos of the masses shown in Fig.13 and derived from Claret's (2004) tracks, it should be noted that we evaluated M for some stars from Heger \& Langer's (2000) tracks with rotation, too. We found that these new M values are somewhat less than previous ones. The difference decreases with decreasing $M$; for instance, the most massive programme supergiant HR 825 reduces its mass from 24 to 22 $M_\odot$, i.e. by 2.0 $M_\odot$, whereas the stars with $M = 10 M_\odot$ (e.g., HR 6144, 7014 and 7770, see Table 2) reduce their masses by 1.2 $M_\odot$ (for $v_0$ = 200 km s$^{-1}$). In all these cases the $M$ changes are within errors of the $M$ determination, so they cannot alter markedly relations presented in Fig.13.

      As noted above, the majority of the observed supergiants are anticipated to be the post-FD objects on the red-blue loops. Fig.13a shows that the supergiants HR 825 and HR 7876 without any visible N enrichment, on the contrary, may be identified as the post-MS objects crossing the Hertzsprung gap for the first time and evolving from slowly rotating MS stars, say, with $v_0$ = 100 km s$^{-1}$ or so. A few other supergiants have the N enhancement less than expected from the FD; these may be post-MS stars suffering from rotationally induced mixing with $v_0$ of about 200 km s$^{-1}$. For the remaining supergiants, Fig.13a suggests that they are post-FD stars having a range of initial rotational velocities of about 50 km s$^{-1}$ to 300 km s$^{-1}$. The bright A2-type supergiant $\alpha$Cyg (Deneb) with the greatest N enrichment in Fig,13, specifically $[N/Fe]$ = 1.02 (Schiller \& Przybilla, 2008), corresponds either to the post-MS or post-FD object with $v_0$ of about 400 km s$^{-1}$. 

      Summarizing, we conclude that our data on the surface nitrogen enrichment in A and F supergiants can be explained within the framework of stellar models with rotational mixing on MS. From the viewpoint of the theoretical predictions, most of the programme supergiants, excluding HR 825 and HR 7876, can be considered to be post-FD stars with a distribution of initial rotational velocities $v_0$ = 50-400 km s$^{-1}$; for the upper end of this range the surface N enrichment should be observable prior to the FD phase.

      It is interesting that our earlier study of the helium abundances in early B-type MS stars with masses $M$ between 4 and 20 $M_\odot$ has led to conclusion that the observed helium enrichment during the MS phase can be explained by rotation with $v_0$ = 250-400 km s$^{-1}$ (Lyubimkov, Rostopchin \& Lambert, 2004). This agreement between results for helium and nitrogen can mean that the initial velocities $v_0$ = 200-400 km s$^{-1}$ are typical for stars with masses $M$~=~4-20~$M_\odot$. 

      Obviously, the key questions now are: What is the distribution of rotational velocities for B-type MS stars, progenitors of the supergiants in question? Is this distribution compatible with that required to reconcile theory with observation according to Fig.13? Do rapidly rotating MS stars show the predicted effects of the rotationally induced mixing? Reviews of the projected rotational velocity $v$sin$i$ for B stars (and other spectral types) are discussed by Bernacca \& Perinotto (1974) and Fukuda (1982) among others. 
The mean $<v$sin$i>$ for B-type MS stars is about 180 km s$^{-1}$. If the rotational axes are randomly inclined to the line of sight, the mean equatorial velocity $<v>$ is $4/\pi$ times $<v$sin$i>$ or 230 km s$^{-1}$. Fukuda fits the $v$sin$i$ distribution with a rotational velocity distribution that is $f(u) = 1/2p$ between the limits (1-p) and (1+p) but vanishes elsewhere and where $u=v/<v>$. He suggests a value of p = 0.4 to 0.5 fits the observed $v$sin$i$ distribution for B main sequence stars, i.e., there is a distribution of rotational velocities from about 110 km s$^{-1}$ to 330 km s$^{-1}$ with a peak at the lower limit, i.e. B-type MS stars with $v_0$ of about 100 km s$^{-1}$ should be plentiful. Given that the majority of the supergiants for HR are plausibly identified as post-FD objects, a velocity distribution from 100 km s$^{-1}$ to 300 km s$^{-1}$ accounts pretty well for the observed N abundances.  

      Interpretations of the N enrichments in supergiants depend in part on an assumption about the initial N abundance. Analyses of B-type MS stars have shown that the stars in the solar neighbourhood have in the mean a composition indistinguishable to within the errors of measurement with that of the Sun. This is true for nitrogen. For example, Venn (1995) who considers the non-LTE N abundance from three main sequence studies (Gies \& Lambert 1992; Kilian 1992, 1994; Cunha \& Lambert 1994) obtains a mean abundance of 7.79$\pm$0.22 for local B-type MS stars, a value essentially that of the Sun (7.83$\pm$0.05, Asplund et al. 2009). More recent representative studies of B-type MS stars include the mean non-LTE N abundances of 7.82$\pm$0.19 (Morel et al. 2006), 7.76$\pm$0.05 (Przybilla et al. 2008) and 7.62$\pm$0.12 (Hunter et al. 2009). In addition, there is good agreement with the N abundance for the Orion nebula with its N abundance of 7.87$\pm$0.09 (Garc\`{i}a-Rojas \& Esteban, 2007). In summary, the initial N abundance of the local supergiants is surely close to the solar value. 

      Disquieting puzzles surround the predictions for rotationally induced mixing: the expectation is that the effects of such mixing in creating a surface N enrichment should be widely seen among rapidly rotating main sequence and evolved B stars. This expectation is not obviously met by Nature's B stars. Evidence for mild N enhancement exists but such enhancements are neither so pervasive nor so severe as expected.  For B-type MS stars, Gies \& Lambert (1992) reported  an asymmetric distribution for their N abundances with a tail to high values and a similar tail to low values of the C abundance; the N/C ratio can be taken as a stronger indicator of mixing than the N abundance by itself. Morel et al. (2009) give N abundances for 20 early-type B main sequence and subgiant stars and discover a population of N-rich slow rotators. The N enhancement is 0.3 dex. Several of these N-rich stars are determined to be slow rotators from `the occurrence of phase-locked UV wind line-profile variations, which can be ascribed to rotational modulation, or from theoretical modelling in the pulsating variables' (i.e., asteroseismology). Our study of helium abundances in early B main sequence stars suggested rotationally induced mixing in stars with $v_0$ of 250-400 km s$^{-1}$ was a common occurrence (Lyubimkov, Rostopchin, \& Lambert 2004). 

      A majority of the abundance studies quoted above have focused for obvious reasons on sharp-lined stars, i.e., stars of low $v$sin$i$. In contrast, Hunter et al. (2009) provide non-LTE N abundances for a large sample of Galactic B stars with $v$sin$i$ values of up to 250 km s$^{-1}$. The primary puzzle arising from this analysis is not so much that the N abundance is independent of $v$sin$i$ but that the more evolved main sequence stars do not show the predicted increase in the N abundance (Hunter et al. 2009, their Fig. 7). Very significantly at the other end of the range of rotational velocities, Morel et al. show that there are (slightly) N-rich B stars with equatorial velocities value too low to have led to a N enhancements according to extant predictions for rotationally induced mixing.

\section{Concluding remarks}

      Our non-LTE nitrogen abundances for 30 A- and F-type supergiants show that a nitrogen enrichment of their atmospheres is nearly universal with 28 of the 30 stars showing enhancements of $[N/H]$ of 0.3 dex or greater. The enhancements are semi-quantitatively consistent with theoretical predictions for post-FD supergiants affected by rotationally induced mixing in main sequence progenitors with rotational velocities having the distribution inferred from measurements of B-type main sequence stars.

      This congruence between observation and theory is, however, not entirely consistent with observations of the composition of B-type MS stars, as we discussed. In short, some rapidly rotating B stars do not have the expected N enhancements and some slow rotators are N enriched. Indeed, there are adherents to the view that local B stars all have a very similar composition (e.g., Przybilla et al. 2008). This unsatisfactory state of affairs shows that observers still need to direct attention to high-precision determinations of the compositions of a LARGE sample of B stars from the main sequence to the supergiant phase with well determined atmospheric stellar parameters including absolute luminosity. In particular, an emphasis should be placed on (i) variables for which brightness and line profile variations will with the tools of asteroseismology provide information on the surface and interior rotational velocities and (ii) stars of high projected rotational velocities where the lines will be broad, shallow, and often blended. Accurate mapping of the surface abundance changes across stellar mass, evolutionary phase, and rotational velocity should provide essential constraints on theoretical ideas including rotationally induced mixing to account for the substantial mixing achieved by B, A, and F-type supergiants ahead of their encountering the first and second dredge-ups. Finally, the excellent work done on the stars in the LMC and SMC should be extended in light of the expectation that the physical processes at play may have a metallicity dependence. 

      Obviously, comparison of the N abundances with the C and O abundance is especially valuable for testing theoretical predictions. For instance, recently Przybilla et al. (2010) considered published data for B-type MS stars and BA-type supergiants, constructed the N/C vs. N/O relation for these stars and compared this empirical relation with the predicted one. They found a confirmation of strong mixing during both in the MS phase and in the supergiant one. We plan to analyse as well the C and O abundances for supergiants of our sample. We hope that a comparison of these homogeneous N, C and O abundances in these stars can provide additional arguments for the mixing during the both evolutionary phases in question.

\section*{Acknowledgments}

      We thank our anonymous reviewer for useful discussion.
      Two of us, LSL and DBP, are grateful to the staff of the Astronomy Department and McDonald Observatory of the University of Texas for hospitality during a visit in October-December 2009. DLL acknowledges the support of the Robert A. Welch Foundation of Houston, Texas through grant F-634.

\appendix
\begin{table*}
\section{}
\bigskip
 \centering
 \begin{minipage}{168mm}
{
\begin{flushleft}
{\bf{Table A1.}} Equivalent widths W(m\AA) and nitrogen abundances $\log\epsilon$(N) for individual NI lines
\end{flushleft}
}
\begin{tabular}{cc@{\,}cc@{\,}cc@{\,}cc@{\,}cc@{\,}cc@{\,}cc@{\,}cc@{\,}cc@{\,}cc@{\,}cc@{\,}cc@{\,}cc@{\,}cc@{\,}cc@{\,}cc@{\,}cc@{\,}cc@{\,}c}

\hline
 Line   & \multicolumn{2}{c}{HR 292} & \multicolumn{2}{c}{HR 825} & \multicolumn{2}{c}{HR 1017} & \multicolumn{2}{c}{HR 1242} & \multicolumn{2}{c}{HR 1740} & \multicolumn{2}{c}{HR 1865} & \multicolumn{2}{c}{HR 2839} & \multicolumn{2}{c}{HR 2874} & \multicolumn{2}{c}{HR 2933}  \\
        &  W & $\log\epsilon$ & W & $\log\epsilon$ & W & $\log\epsilon$ & W & $\log\epsilon$ & W & $\log\epsilon$ & W & $\log\epsilon$ & W & $\log\epsilon$ & W & $\log\epsilon$ & W & $\log\epsilon$ \\
\hline                                                                                                                      
7423.64 &  -  &  -   &   78 & 7.94 &   -  & -    &  -  &  -   &   -  &  -   &   -  &  -   &   -  &  -   &   -  &  -   &  -  &  -   \\
7442.29 &  94 & 8.51 &  134 & 7.90 &   80 & 8.48 &  83 & 8.33 &  110 & 8.14 &  156 & 8.69 &   -  &  -   &   -  &  -   &  68 & 8.34 \\
7468.31 & 116 & 8.49 &  183 & 7.87 &   96 & 8.40 & 110 & 8.35 &  135 & 8.09 &  192 & 8.71 &   -  &  -   &   -  &  -   &  89 & 8.34 \\
8184.86 & 102 & 8.47 &  167 & 7.87 &   82 & 8.38 &  95 & 8.33 &  123 & 8.12 &  180 & 8.73 &   -  &  -   &   -  &  -   &  85 & 8.40 \\
8188.01 & 112 & 8.56 &  170 & 7.88 &    - &   -  & 103 & 8.40 &  133 & 8.18 &   -  &  -   &   -  &  -   &   -  &  -   &  -  &  -   \\
8200.35 &  -  &  -   &   -  &  -   &    - &   -  &  -  &  -   &   -  &   -  &   -  &  -   &   39 & 8.47 &   -  &  -   &  -  &  -   \\
8210.71 &  -  &  -   &   -  &  -   &    - &   -  &  -  &  -   &   -  &   -  &   -  &  -   &   92 & 8.38 &   -  &  -   &  -  &  -   \\
8216.33 &  -  &  -   &   -  &  -   &    - &   -  &  -  &  -   &   -  &   -  &   -  &  -   &  174 & 8.22 &   -  &  -   & 141 & 8.26 \\
8223.12 &  -  &  -   &   -  &  -   &    - &   -  &  -  &  -   &   -  &   -  &  216 & 8.62 &  124 & 8.24 &   -  &  -   & 117 & 8.47 \\
8242.38 &  -  &  -   &  167 & 7.76 &    - &   -  &  -  &  -   &   -  &   -  &   -  &  -   &   -  &  -   &   -  &  -   &  -  &  -   \\
8680.28 &  -  &  -   & (502)& 7.85 &    - &   -  &   - &   -  & (311)& 8.09 & (366)& 8.73 &   -  &  -   & (496)& 8.16 &  -  &  -   \\
8683.40 &  -  &  -   & (376)& 7.85 & (172)& 8.37 & 190 & 8.33 & (246)& 8.05 & (315)& 8.74 & (214)& 8.46 & (404)& 8.17 & 167 & 8.41 \\
8686.14 &  -  &  -   & (257)& 7.98 &    - &   -  &  -  &  -   & (183)& 8.15 & (245)& 8.78 & (162)& 8.49 & (310)& 8.30 & 131 & 8.55 \\
8703.24 &  -  &  -   &  239 & 7.94 &    - &   -  & 131 & 8.34 &   -  &  -   &  246 & 8.79 &  155 & 8.45 &  307 & 8.30 & 115 & 8.43 \\
8711.70 &  -  &  -   &  264 & 7.92 &    - &   -  & 142 & 8.33 &  196 & 8.13 &  254 & 8.74 &  164 & 8.42 &  319 & 8.24 & 126 & 8.42 \\
8718.83 & 124 & 8.40 &  222 & 7.91 &    - &   -  & 122 & 8.28 &  176 & 8.14 &  232 & 8.73 &  148 & 8.40 &  293 & 8.27 & 113 & 8.42 \\
8728.90 &  -  &  -   &   -  &  -   &    - &   -  &  54 & 8.42 &   -  &  -   &   97 & 8.65 &   59 & 8.37 &   -  &   -  &  -  &  -   \\
\\
\hline
 Line   & \multicolumn{2}{c}{HR 3183} & \multicolumn{2}{c}{HR 3291} & \multicolumn{2}{c}{HR 6081} & \multicolumn{2}{c}{HR 6144} & \multicolumn{2}{c}{HR 6978} & \multicolumn{2}{c}{HR 7014} & \multicolumn{2}{c}{HR 7387} & \multicolumn{2}{c}{HR 7770} & \multicolumn{2}{c}{HR 7796}  \\
        &  W & $\log\epsilon$ & W & $\log\epsilon$ & W & $\log\epsilon$ & W & $\log\epsilon$ & W & $\log\epsilon$ & W & $\log\epsilon$ & W & $\log\epsilon$ & W & $\log\epsilon$ & W & $\log\epsilon$ \\
\hline
7423.64 &   -  &  -   &   -  &  -   &   -  &  -   &  127 & 8.69 &   -  &  -   &   -  &  -   &   -  &  -   &   -  &  -   &   -  &  -   \\
7442.29 &   90 & 8.12 &  130 & 8.58 &  121 & 8.33 &   -  &  -   &   -  &  -   &   96 & 8.33 &  130 & 8.56 &   51 & 8.19 &   -  &  -   \\
7468.31 &  116 & 8.10 &  169 & 8.62 &  145 & 8.30 &  192 & 8.52 &   -  &  -   &  121 & 8.30 &  169 & 8.59 &   69 & 8.18 &   67 & 8.27 \\
8184.86 &   -  &  -   &   -  &  -   &  136 & 8.34 &  176 & 8.54 &   -  &  -   &  130 & 8.45 &  156 & 8.60 &   -  &  -   &   69 & 8.39 \\
8188.01 &   -  &  -   &   -  &  -   &   -  &  -   &   -  &  -   &   -  &  -   &   -  &  -   &  172 & 8.70 &   -  &  -   &   -  &  -   \\
8200.35 &   -  &  -   &   -  &  -   &   -  &  -   &   -  &  -   &   -  &  -   &   -  &  -   &   -  &  -   &   -  &  -   &   -  &  -   \\
8210.71 &   -  &  -   &   -  &  -   &   -  &  -   &   -  &  -   &   -  &  -   &  107 & 8.47 &   -  &  -   &   -  &  -   &   -  &  -   \\
8216.33 &   -  &  -   &   -  &  -   &   -  &  -   &   -  &  -   &   98 & 8.19 &   -  &  -   &   -  &  -   &  118 & 8.08 &  108 & 8.17 \\
8223.12 &   -  &  -   &  215 & 8.73 &  158 & 8.20 &  -   &  -   &  -   & -    &  137 & 8.26 &  178 & 8.49 &   74 & 8.13 &   -  &  -   \\
8242.38 &   -  &  -   &   -  &  -   &   -  &  -   &  -   &  -   &  -   & -    &  -   & -    &  -   & -    &   81 & 8.18 &   -  &  -   \\
8680.28 & (277)& 8.16 &   -  &  -   & (289)& 8.30 &  -   &  -   &  -   & -    &  -   & -    &  -   & -    &   -  &  -   &   -  &  -   \\
8683.40 &   -  &  -   & (288)& 8.66 & (234)& 8.24 & (345)& 8.60 & (101)& 8.19 & (219)& 8.28 & (292)& 8.63 & (125)& 8.11 & (124)& 8.25 \\
8686.14 & (147)& 8.12 & (211)& 8.66 & (188)& 8.36 & (258)& 8.61 &   -  &  -   & (174)& 8.44 & (230)& 8.72 &   -  &  -   &   -  &  -   \\
8703.24 &  150 & 8.14 &  211 & 8.67 &  181 & 8.33 &  256 & 8.61 &   71 & 8.32 &  160 & 8.37 &  209 & 8.62 &   73 & 8.10 &   -  &  -   \\
8711.70 &  160 & 8.11 &  230 & 8.68 &  188 & 8.28 &  279 & 8.62 &   -  &  -   &  164 & 8.30 &  217 & 8.57 &   -  &  -   &   -  &  -   \\
8718.83 &  143 & 8.11 &  198 & 8.61 &  169 & 8.27 &  240 & 8.55 &   58 & 8.19 &  156 & 8.35 &  199 & 8.58 &   76 & 8.14 &   -  &  -   \\
8728.90 &   -  &  -   &   -  &  -   &   72 & 8.34 &   -  &  -   &   -  &  -   &   -  &  -   &   -  &  -   &   -  &  -   &   -  &  -   \\
\\
\hline
 Line   & \multicolumn{2}{c}{HR 7823} & \multicolumn{2}{c}{HR 7834} & \multicolumn{2}{c}{HR 7847} & \multicolumn{2}{c}{HR 7876} \\
        &  W & $\log\epsilon$ & W & $\log\epsilon$ & W & $\log\epsilon$ & W & $\log\epsilon$ \\
\hline
7423.64 &   -  &  -   &   -  &  -   &   -  &  -   &   -  &  -   \\
7442.29 &  107 & 8.51 &   -  &  -   &   76 & 8.39 &   -  &  -   \\
7468.31 &  122 & 8.41 &   -  &  -   &   98 & 8.38 &   62 & 7.77 \\
8184.86 &  102 & 8.37 &   -  &  -   &   90 & 8.41 &   58 & 7.83 \\
8188.01 &   -  &  -   &   -  &  -   &   -  &  -   &   -  &  -   \\
8200.35 &   -  &  -   &   -  &  -   &   -  &  -   &   -  &  -   \\
8210.71 &   -  &  -   &   -  &  -   &   -  &  -   &   -  &  -   \\
8216.33 &  202 & 8.37 &  116 & 8.15 &  153 & 8.27 &  115 & 7.65 \\
8223.12 &  143 & 8.42 &   -  &  -   &   95 & 8.25 &   -  &  -   \\
8242.38 &  130 & 8.32 &   -  &  -   &   96 & 8.25 &   -  &  -   \\
8680.28 &   -  &  -   &   -  &  -   &   -  &  -   &   -  &  -   \\
8683.40 & (207)& 8.36 & (123)& 8.18 & (176)& 8.38 & (131)& 7.73 \\
8686.14 & (172)& 8.56 &   -  &  -   &   -  &  -   &   -  &  -   \\
8703.24 &  171 & 8.56 &   83 & 8.26 &  115 & 8.38 &   -  &  -   \\
8711.70 &   -  &  -   &   86 & 8.19 &   -  &  -   &   -  &  -   \\
8718.83 &  149 & 8.44 &   -  &  -   &  107 & 8.33 &   -  &  -   \\
8728.90 &   -  &  -   &   -  &  -   &   -  &  -   &   -  &  -   \\
\hline

\end{tabular}   
\end{minipage}  
\end{table*}

\bsp            
                
\label{lastpage}
                
\end{document}